# Bypass Fraud Detection:
## Artificial Intelligence Approach

Dr. Ibrahim Ighneiwa
Department of Electrical and Electronics Engineering
Faculty of Engineering, University, of Benghazi
Benghazi, Libya
Ibrahim.ighneiwa@uob.edu.ly

Hussamedin S. Mohamed
Department of Electrical and Electronics Engineering
Faculty of Engineering, University, of Benghazi
Benghazi, Libya
Hsbargathy@gmail.com

*Abstract*—Telecom companies are severely damaged by bypass fraud (SIMboxing). However, There is a shortage of published research to tackle this problem. The traditional method of Test Call Generators (TCG) is easily overcome by fraudsters and the need for more sophisticated ways is inevitable. In this work, we are developing intelligent algorithms that mine a huge amount of mobile operator's data and detect the SIMs that used to bypass international calls. This method will make it hard for fraudsters to generate revenue and hinder their network. Also by reducing fraudulent activities, quality of service can be increased as well as customer satisfaction. Our technique has been tested and evaluated on real world mobile operator data, and it proved to be very efficient.

## I. INTRODUCTION

Imagine you get an unknown call from a local number, you pick up the phone and it's from a friend or a family living abroad, it does feel strange that you would receive an international call from a local number; this is basically the bypass fraud, or SIMboxing. In this case, the international calls are transferred over the Internet to a cellular device that injects them back into the cellular network through SIM boxes with multiple low-cost prepaid SIM cards. As a result, the calls turn local at the destination network and the fraudsters who set up these boxes pay only local rates for mobile operators after charging international rates from the source. The person calling will pay the whole call termination fee, but it would not be collected by his/her local operator; this fee will go to the fraudster responsible for the SIM box and other companies that are responsible for directing the calls from source to destination.

Most people think of fraudsters as hackers, but this is not always the case, most of the time they are business people, they perceive the fraudulent activity as a business opportunity, where they are willing to invest time and money to gain benefit. So, we believe that fraudsters could be defeated by simply putting them out of business.

This work was supported in part by Almadar Aljadeed Mobile Phone Company and in part by the University of Benghazi, Benghazi, Libya

In this work, a detection system is designed and tested in cooperation with the Tier 1 mobile operator in Libya (Almadar Aljadeed Mobile Phone Company). The system is utilizing artificial intelligence techniques to detect whether a SIM card is used by a normal customer or by a fraudster.

## II. IMPACT OF FRAUD

Telecommunications, which have become a necessity worldwide, became a target for fraudsters who are making a lot of money out of illegally accessing communication networks and using it to make huge profits, by selling services at much lower prices than their original prices. According to a survey by the Communication Fraud Control Association (CFCA) [1], the mobile telecom industry lost more than 38 billion dollars in 2015 alone due to telecom fraud. Besides those big losses, telecom fraud causes other indirect losses to mobile operators, like: decrease in quality of service, deny of service and network congestion.

Bypass fraud costs telecom companies 6 billion annually and ranked the $2^{nd}$ most costly fraud worldwide. Fig. 1 shows the top 3 fraud types with their annual losses. The numbers are huge, since major mobile operators in Libya are owned by the state, the revenue obtained by these major companies could help in Libya's economic growth, and anything that would affect it would degrade the country's GDP.

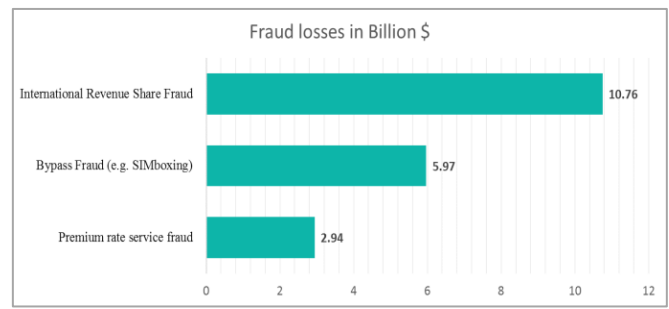

Fig. 1. CFCA 2015 survey, Top 3 fraud losses globally

## III. TYPES OF FRAUD

Phone systems fraud is not just due to phone theft or hacking, It is much more than that. Yelland [2] lists seven of them. The top three types that cause a significant loss are:

### A. International Revenue Share Fraud (IRSF)

IRSF is the largest contributor to the overall fraud losses according to CFCA. It is when a Fraudester makes an agreement with a local carrier in high cost destination to share profit for increasing traffic, then the fraudster hacks into any organization's public branch exchange (PBX) and gets illegal access to generate calls. After that, the fraudster generates high traffic calls to high cost destinations and gets revenue from the sharing agreements. [4].

### B. Premium Rate Service Fraud

It must have happened to you; that is receiving a text message saying that you won a big prize and all what you need to do is to call a certain number. If you called that number you have been tricked to use some premium rate service. Premium rate service is an agreement between some service provider and telecom companies to share revenues generated by traffic to the premium service number, it is used in TV shows and contests and entertainment services. The fraudsters try to stimulate the costumers by giving them a missed call or a message, then make money from share of the call back revenues [4].

### C. Bypass fraud

To explain how bypass fraud is committed, firstly we describe the legitimate way for international calls. Let's assume that caller A and caller B live in different countries. Caller A makes a call to caller B over the mobile operator. The mobile operator of country A takes the call and send it through his international gate to a transient operator. The transient operator then routes the call through voice over IP (VoIP) to the country B mobile operator and pays a toll. After that, the mobile operator of country B terminates the call through his network to caller B. Fig. 2 shows the legitimate route.

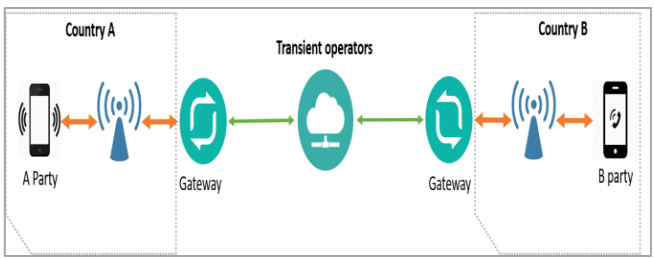

Fig. 2. The legitimate route of international call

In bypass fraud, the transient operator route the call through a SIMbox placed in country B using VoIP, the SIMbox then reroute the call through country B mobile operator and pay for just the local call. Fig. 3 shows the bypass fraud route.

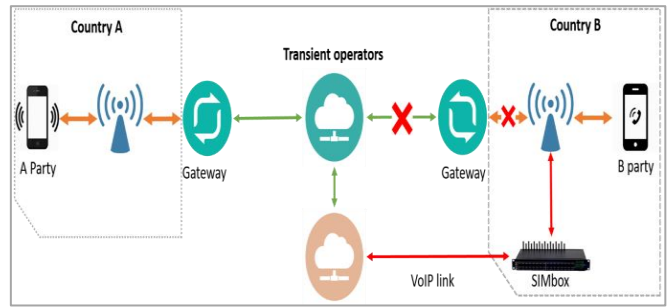

Fig. 3. The bypass fraud route of international call

The incentive here is the toll charge by country B mobile operator is much higher than the local call fee, so the bypass will be financially viable.

Bypass fraud is committed when fraudsters install SIMboxes with multiple low-cost, prepaid SIM cards. SIMbox equipments includes SIM slots, antennas. The SIMbox is connected to the internet through Ethernet port. Fig. 4 shows a SIMbox and its components.

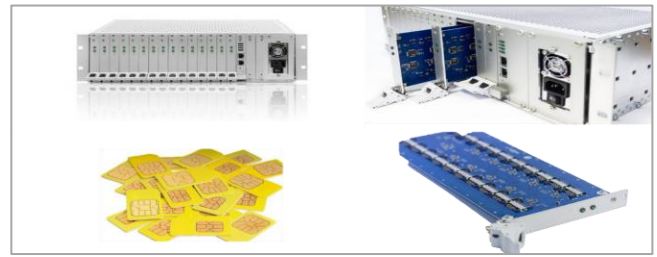

Fig. 4. A SIMbox and its components

## IV. BATTLING BYPASS FRAUD

The major methods used today for battling fraudsters are:

### A. Test Call Generation (TCGs)

Test call generation is used as an active method to detect bypass fraud, where operators test different international routes to their network and see whether calls go through legitimate routes or SIMbox routes. This method detects fraud with no false positive (that is when a normal user assumed to be a SIMbox). However, this method is probabilistic in nature and costly in terms of the need to test huge number of international routes. Also, fraudsters use tricks to avoid test call detection as we will see in the anti-spam method.

### B. Fraud Management Systems (FMS)

Fraud management systems use measures to detect the abnormal usage of SIM cards. FMS analyze Call Details Record (CDR) data to make usage profiling that distinguishes normal users from SIMboxes.

### C. SIM Card Distribution Control (SDC)

SIM cards are vital in the bypass cycle, and fraudsters must maintain an adequate supply of SIM cards to be in business. However, SIM card distribution control will make this process difficult. Requiring government IDs and limiting the number of SIM cards per ID will prevent fraudsters from obtaining a large number of SIM cards to install in their SIMboxes.

## V. HOW FRAUDSTERS AVOID DETECTION

Fraudsters and anti-fraud are in eternal battle, every time detection technology improves, fraudsters are developing their methods to avoid detection and increase profit. This section describes different methods used by fraudsters to avoid SIM blocking.

### A. Anti-Spam (Test Call Detection)

One of the effective methods to detect SIMs used in SIMboxes is generating test calls (TCG) using different routes to known local network numbers. The incoming call will appear weather it is coming from a local number or from an international number; if it was coming from a local number then it must be associated with some SIM card used in a SIMbox and easily processed by the fraud department. However, the fraudsters analyze the voice call traffic coming toward their SIMboxes and based on usage and other patterns they could determine whether the calls were real subscriber calls or they were originated from a TCG system. They coud then either block the test calls and prevent them from reaching the SIM box, to begin with, or reroute the calls to a legitimate route so as to avoid detection.

### B. Human Behavior Simulation (HBS)

According to the literature [3], some features can be used to identify SIMbox fraud, for example: 1. The SIMbox is not moving. 2. Most calls are outgoing calls. 3. No usage of network services like SMS, GPRS. and others. However, Smart SIMboxes are designed to mimic the behavior of normal customers by using Human Behavior Simulation (HBS). This technique makes detection of fraudsters very difficult if no advanced detection algorithms were used. HBS encompasses the following:

#### 1) SIM Migration (Movability)

Fraudsters are deploying many gateways in different locations, for example, one in the city center and another in a shopping mall or some other crowded place and once in a while they swap the SIM cards between the gateways, so it would look like that the user is moving. The swapping operation could be done manually or automatically using software.

#### 2) SIM Rotation

SIMboxes can be detected easily if fraudsters operate their SIMs around the hour excessively, so they limit their usage by rotation of the SIMs as workers shifts. This will make SIMs operate in limited hours a day, which simulates the behavior of ordinary customers.

#### 3) Usage of Other Network Services

Most of the SIMboxes are using just voice services and that makes them vulnerable to detection. In order to mitigate this issue smart SIMboxes are making calls and sending SMS to each other. Also, sometimes they use some internet services provided by the network operator.

#### 4) Family Lists

Traditional SIMboxes just reroute the call from VoIP to the GSM network, so they make calls to large numbers of different network customers. A smart way to avoid this is by using family lists, where each SIM is assigned to reroute calls to a specific list of numbers. This leads to escaping the trap of large different numbers detection.

To summarize, Human Behavior Simulation (HBS) makes dealing with bypass fraud harder and harder and time-consuming. Advanced measures must be taken to tackle this problem. In this work, intelligent machine learning algorithms were used to detect the bypass fraud by analyzing huge CDR data.

## VI. LITERATURE REVIEW

Even though the telecommunication industry suffers major losses due to fraud, there is no comprehensive published research on this area mainly due to lack of publicly available data to perform experiments on. On the other hand, any broad research published publicly about fraud detection methods will be utilized by fraudsters to evade detection.

This section presents the literature of the available research published. Most research investigated CDR analysis combined with machine learning algorithms to detect fraudulent SIMboxes [4] [5] [3], and few others used Audio analysis [6].

In [4], 234,324 calls made by 6415 subscribers from one Cell-ID during two months were analyzed. The dataset consisted of 2126 fraud subscribers and 4289 normal subscribers which are equivalent to two thirds of legitimate subscribers and one third of fraudulent SIMboxes. The researchers extracted 9 features, like Total Calls, Total Number Called, Total Minutes and Average Minutes, etc. Then they used the extracted features to train an Artificial Neural Network (ANN) classifier, where three architectures of neural networks were considered and three hidden layers; 5, 9 and 18 hidden nodes in each layer. They found that the best architecture was when two hidden layers were used, each having five hidden neurons, with a learning rate of 0.6 and a momentum term of 0.3. The accuracy reached 98.7% with just 20 accounts were wrongly classified as false positive.

In [5], researchers extended their previous work in [4] to design and compare two classifiers; Support Vector Machine (SVM) classifier and ANN classifier. Same dataset and features were used. They found that SVM has better accuracy compared to ANN. SVM gave 99.06% accuracy while ANN model gave 98.69% accuracy. In addition, the SVM training time was found to be three times less than the time consumed by the ANN training time.

Another more broad research was conducted by [3]. In contrast with [4] and [5], where data was constrained to only one Cell-ID, a larger dataset was used with accounts distributed nationwide. They analyzed CDRs form main cellular operator in the United States (AT&T). The dataset contains CDRs of 93000 legitimate accounts and 500 of fraudulent accounts. For training the classifier they split the dataset to two-thirds for training and one third for testing. Using International Mobile Equipment Identity (IMEI) as a device identifier other than the subscriber identifier they computed 48 features characterizing patterns of legitimate and fraudulent IMEIs.

They observed that fraudulent SIMboxes have common patterns as the following:

1) High number of Int. mobile subscriber identity number (IMSIs) per IMEI.
2) Static physical location.
3) Large number of international phone calls.
4) Large volume of outgoing calls generated compared to incoming calls.

For classification, they used a combination of decision tree classifiers (alternating decision tree, functional tree, and random forest). The classification rule was the linear combination of their results. The accuracy of the classification rule was 99.95%. with lowest false positive achieved by the random forest.

On the other hand, instead of using CDR analysis, [6] used real-time call audio analysis to detect fraudulent calls. They designed a system that relies on the raw voice data received by the tower during a call to distinguish errors in GSM transmission from the distinct audio distortions caused by delivering the call over a VoIP. They used fast signal processing techniques to identify whether individual calls are likely made by a SIMbox and then to develop profiles of SIM cards. Their resultant system was able to detect 87% of real SIMbox calls in only 30 seconds of audio with no false positives. Their system promises a real-time detection capability if it were deployed on network towers and embedded on the operators Base Transceiver Station (BTS).

VII. Designing the detectin system

Designing the detection system was done in four stages:
- Data Collection and Cleaning: In cooperation with Almadar Aljadeed Mobile Phone Company, fully anonymized Call Detailed Records (CDR) were obtained and utilized. The CDR's sensitive data like user numbers or identities were obfuscated and hashed by the mobile operator. CDR data was processed in order to remove missing values, duplicate information and useless fields (here we used Pathon filters).
- Feature Extraction and Engineering: To prepare the data for the machine learning model, informative features were extracted. In this stage features for each SIM were extracted. Also, features had been engineered in order to get the best features for training the model. Feature engineering includes feature selection and dimensionality reduction in generating more useful features.
- Model Training: The features extracted in the previous stage were used to train machine learning models. In this stage, different models were trained to detect SIMs that were used in SIMboxes. SVM and Decision trees (random forest algorithm) had been used as supervised learning algorithms, and since the labeled data is scarce, we used unsupervised learning algorithms to cluster the SIMs in order to get insights of how we could improve the designed algorithm.
- Model Evaluation and Testing: The final stage was testing and comparing the performance of designed algorithms in terms of accuracy and precision and that showed that our system is very efficient. We tested our system on real time data on site, where we took our system to our sponsor (Almadar Aljadeed Mobile Phone Company) headquarters and our tests showed the high efficiency of our bypass fraud detection system.

Fearing that fraudesters might use information published here to their advantage, affected how much details we could reveal, but, our data and detailed work techniques would be available for all of those concerned in our presentation.

VIII. Conclusions

We have studied the problem of bypass fraud (SIMboxing) and found that although bypass fraud has been damaging telecom companies severely, there has been a shortage of published research to study it and solve it. So our work has been two fold, one is to increase awareness about this big problem and to show that conventional techniques could not be used to solve it. We utilized some intelligent techniques to effectively detect SIMboxing fraud and prevent it from affecting telecomm companies not only when it comes to revenues but also when it comes to denial of service, quality of service, and communications network congestion.